# Identifying the Main Factors of Iran's Economic Growth using Growth Accounting Framework

Mohammadreza Mahmoudi

*Department of Economics, Northern Illinois University, Dekalb, USA.*

Email: mmahmoudi@niu.edu

**Correspondence**

Mohammadreza Mahmoudi, *Northern Illinois University, Dekalb, USA.*

Email: mmahmoudi@niu.edu

## Abstract

This paper aims to present empirical analysis of Iranian economic growth from 1950 to 2018 using data from the World Bank, Madison Data Bank, Statistical Center of Iran, and Central Bank of Iran. The results show that Gross Domestic Product (GDP) per capital increased by 2 percent annually during this time, however this indicator has had a huge fluctuation over time. In addition, the economic growth of Iran and oil revenue have close relationship with each other. In fact, whenever oil crises happen, great fluctuation in growth rate and other indicators happened subsequently. Even though the shares of other sectors like industry and services in GDP have increased over time, the oil sector still plays a key role in the economic growth of Iran. Moreover, growth accounting analysis shows contribution of capital plays a significant role in economic growth of Iran. Furthermore, based on growth accounting framework the steady state of effective capital is 4.27 for Iran's economy.

Keywords: Growth Economics, Solow Model, Growth Accounting, Iran's Economy

## 1. Introduction

Iran is a country in the Middle East with over 81 million inhabitants and is the world's 18th most populous country. Also, it is the 17th largest country in the world in terms of area. Moreover, Iran has the world's second largest natural gas reserves and the fourth largest proven crude oil reserves (World Development Indicators, 2019).

In the Twentieth Century we see a great change in the economy of Iran. At the beginning of the century, Iran is an undeveloped country, while in the latter half of the 20th century the path of Iran's economic development changed dramatically (Esfahani and Pesaran, 2008). During these years GDP per capita increased by about seven times. Other changes in different indicators are summarized in Table 1.

*Table 1. Change in the quantity of some indicators during the 20th century*

| Indicator | 1900 (Rough Estimate) | 2018 |
|---|---|---|
| **Population (millions)** | 8.6 | 80.1 |
| **Rank in the World** | 20 | 18 |
| **Out of ….** | 228 | 198 |
| **Urban population (% of total)** | 27% | 73% |
| **GDP per capita (constant 2010 US$)** | $1,000 | $6947 |
| **Rank in the World** | 44 | 93 |
| **Out of ….** | 68 countries for which data is available | 184 |
| **Agriculture, forestry, and fishing, value added (% of GDP)** | 65% | 10% |
| **Life Expectancy (Total)** | <40? | 76 |
| **Rank in the World** | ? | 62 |
| **Literacy Rate (Population 15+)** | <5? | 85% |

Source: Maddison Project Database (MPD) 2018; World Bank, World Development Indicators 2018; Statistical Center of Iran

Table 1 does not show the fluctuations that occurred during the 20th century, therefore Table 2 shows the mean and standard deviation of some critical indicators in each decade following 1960.

*Table 2. The statistical features of some critical indicators*

<table>
<tr><th></th><th>Average</th><th>Standard Deviation</th></tr>
<tr><td>Real GDP growth rate (annual %)</td><td>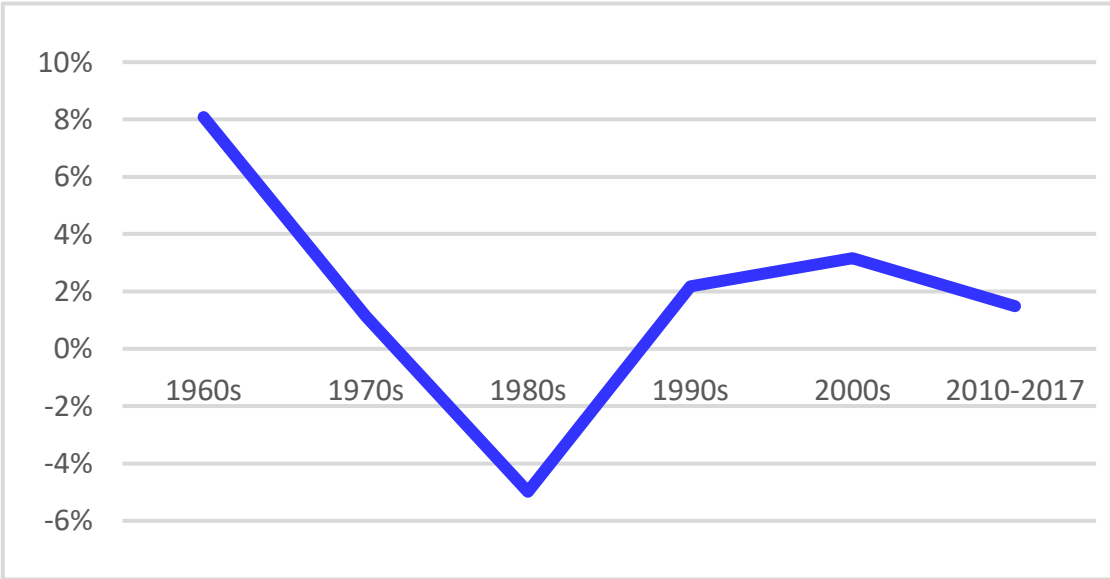
</td><td>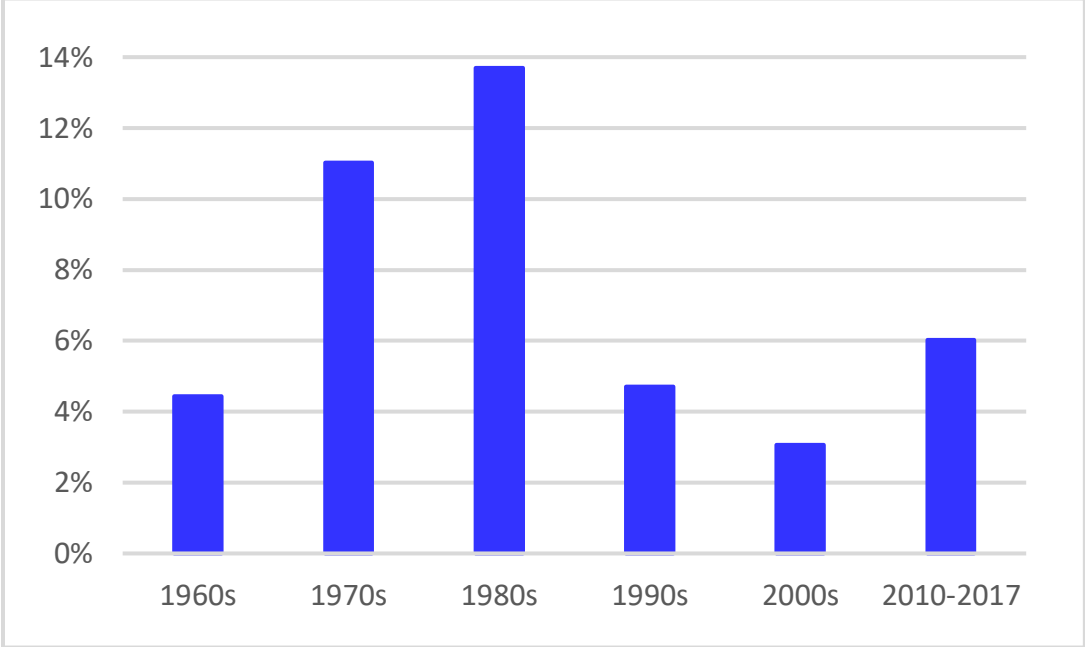
</td></tr>
<tr><td>Population growth rate (annual %)</td><td>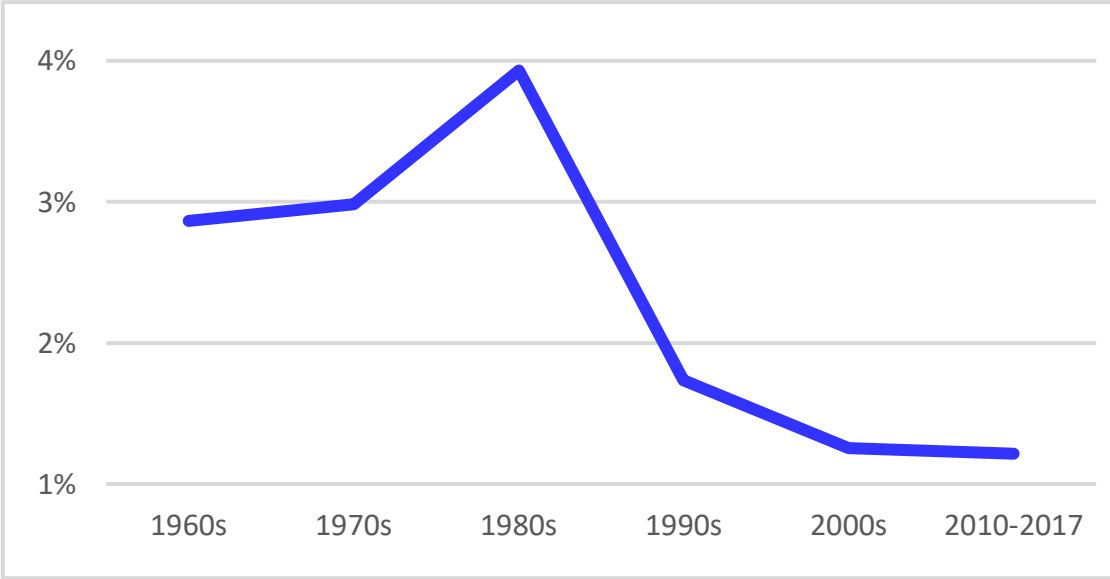
</td><td>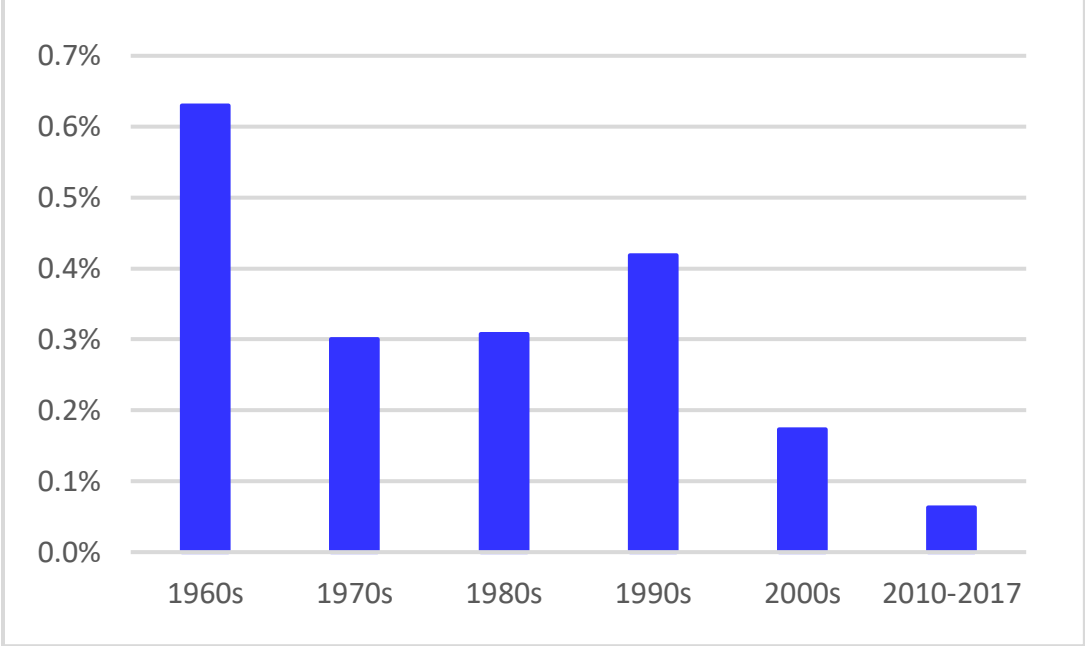
</td></tr>
<tr><td>Inflation</td><td>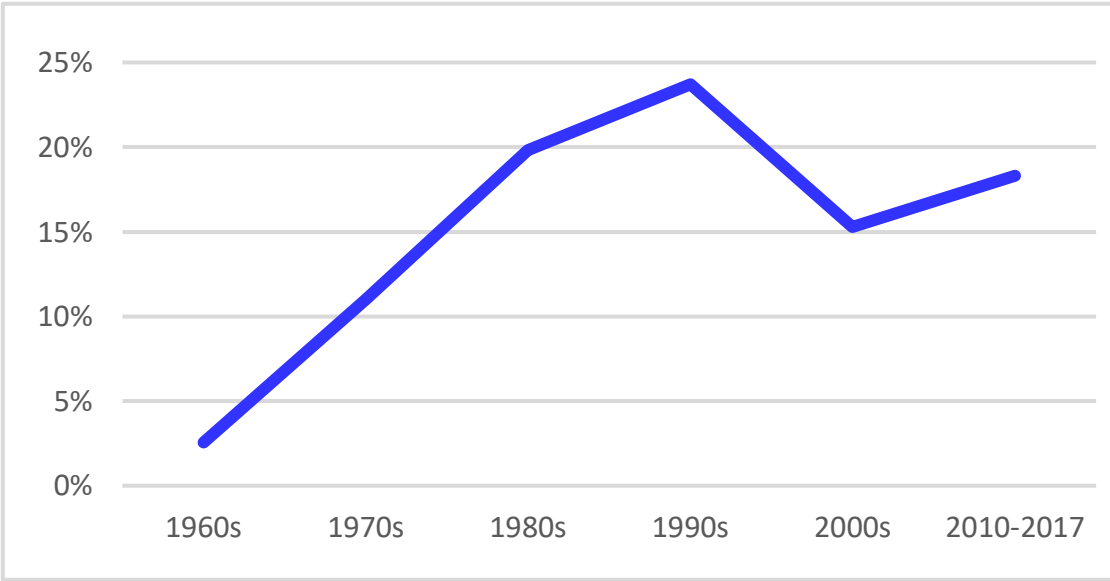
</td><td>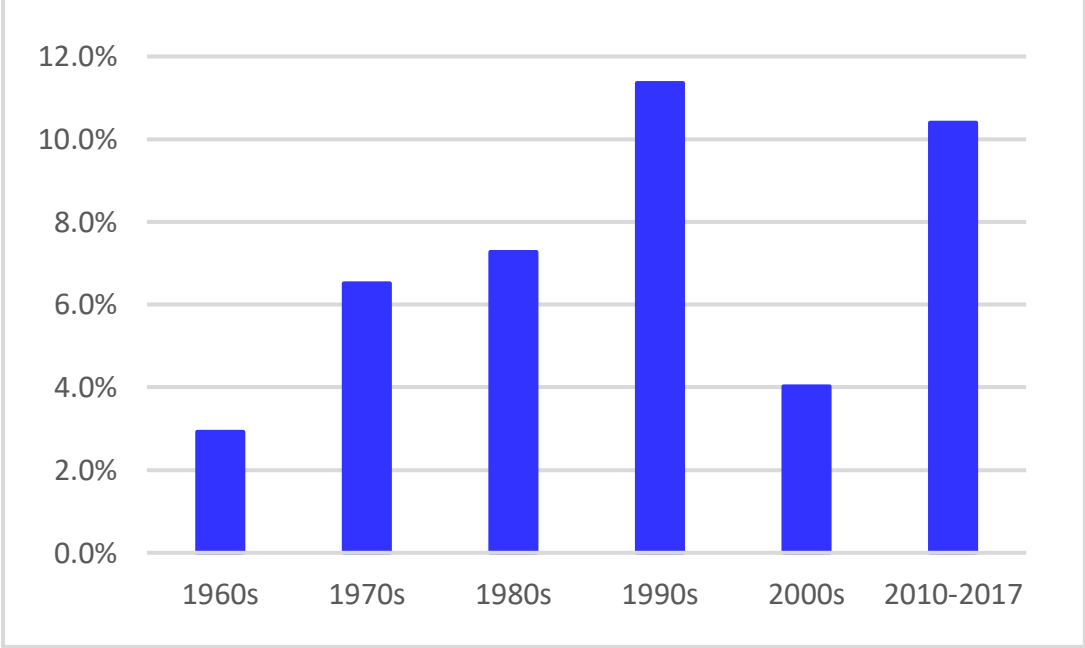
</td></tr>
</table>

Source: World Bank, World Development Indicators 2018

As described in Table 2, the best decade in terms of average GDP per capita is the 1970s, however, this decade had the highest value of standard deviation for GDP per capita. The highest quantity of average rate of GDP growth was reported in 1960s. In comparison, the minimum quantity of average rate of GDP growth and the maximum standard deviation of this indicator belong to the 1980s. In addition, the highest quantities of average and standard deviation of inflation were reported in the 1980s.

The main goal of this paper is to detect and explain the sources of Iranian economic growth over the contemporary era. To this end, this paper is divided into two parts. In the first part, we summarize some of the main papers' conclusions, which study economic growth using growth accounting framework. In the second part, we describe the economic growth of Iran over time. Finally, based on the growth accounting framework, the reasons behind Iran's economic growth rate are discussed.

## 2. Literature Review

Growth accounting is used to break down how specific factors contribute to economic growth. Estimation of various important variables and testing various propositions or alternative specifications of the growth theory are main goals of growth accounting. In fact growth accounting helps to get better understanding regarding economic growth (Acemoglu, 2010).

(Solow, 1957)introduced the growth accounting approach for the first time. (Gollop et al., 1987) developed this approach by emphasizing the necessity of the competitive market. The results show that the physical and human capital underestimation leads to overestimation of the contribution of technology to economic growth. Also, growth regressions, which developed by (Barro, 1991) become a very commonly used technique in recent years. For various studies of growth regressions, see (Durlauf, 1996); (Durlauf et al., 2005); and (Wooldridge, 2002). Moreover, the augmented Solow model with human capital is a generalization of the model presented in (Mankiw et al., 1992). (Jones, 2001)emphasized the "people produce ideas" channel as a channel to generate economic growth, while (Hansen and Prescott, 2002) introduces a structural transformation from agriculture to manufacturing as an economic growth channel. (Caselli, 2005) discussed how to correct differences in the quality of physical and human capital across countries. (Jones, 2016) provide an encyclopedia of the fundamental facts of economic growth based on growth theories and update the facts with the newest available data. (Atkeson et al., 2019) study optimal policy in the presence of several creative channels. (Garcia-Macia et al., 2019) analyze the types of innovation behind US growth in recent decades using LBD data.

In addition, there are several papers which study the growth economics of Iran. (Jbili et al., 2004) present a comprehensive study regarding Iran's economic growth. (Esfahani and Pesaran, 2008) study the economic transformation of Iran in a global context through the Twentieth Century. Their results show that Iran's economy transformed significantly during the Twentieth Century, however, the initial conditions and the evolution of domestic institutions and resources and relations with the rest of the world played critical roles in that transformation processing. (Mojaver, 2009) uses a standard measure of Total Factor Productivity (TFP) to analyze the differences between the economic performance of Iran in the two decades before and after the Islamic revolution in 1979. The results show that TFP growth rates explain 30 percent of the difference in economic performance.

## 3. Analyzing Iran's Economic Growth over the contemporary era

In order to get a comprehensive picture regarding economic growth in Iran we focus on GDP per capita from 1950 to 2018. As shown in Figure 1, GDP per capita fluctuated drastically over time, especially after the Islamic Revolution in 1979.

*Figure 1.GDP per person in Iran*

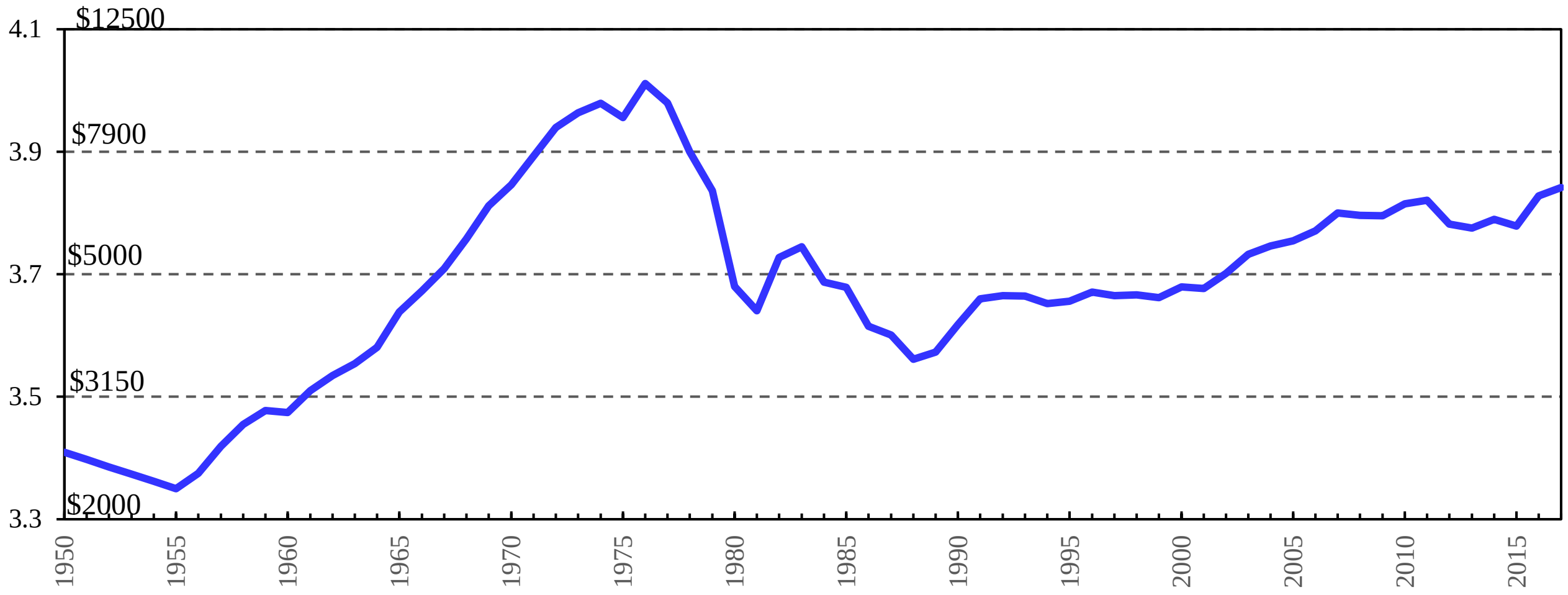

Source: Maddison Project Database (MPD) 2018; World Bank, World Development Indicators 2018

The maximum quantity of GDP per capita is $10,266 which was reported in 1976 (three years after 1973 oil crisis and three years before the Islamic Revolution) and the minimum quantity is $2,238 which was reported in 1955 (six years before the Industrialization period). Although the annual GDP per capita growth rate is around 2 percent on average, this indicator had huge fluctuations over this time. Knowing the reasons behind these variations, we split the contemporary era of Iran into several important periods and explain related economic and political situations in each period specifically.

- **Period 1: 1925-1941**

Before 1950, specifically from 1925 to 1941, Reza Shah Pahlavi, the first leader of modern government in Iran, improved infrastructure profoundly. During this period, a lot of roads, hospitals, and schools were constructed so that the effect of this period on Iranian economy is inevitable (Pesaran, 1997)

- **Period 2: 1961-1972**

This period is famous as the golden age of the Iranian economy, because only at these years Iran experienced strong and stable long-term economic growth. During this period, GDP per capita grew by an average annual rate of around 9 percent with the annual population growth rate of around 3 percent and average annual inflation rate of around 2 percent. As we mentioned before, infrastructure improved significantly in previous years. Moreover, due to a rise in oil revenues, the government invested a lot of money to develop oil and non-oil industries and promote manufacturing. Indeed, the role of professional economists in improving economic policies is undeniable (Esfahani and Pesaran, 2009, Esfahani and Squire, 2007).

- **Period 3:1972-1978**

In this period, the oil revenues reached their highest level around 1973, but the major challenge for the government was managing them properly (Pesaran, 1997).Expansionary fiscal and monetary policy led to high and rising inflation in the mid-1970s. The government's attempt to control inflation led to increase economic instability and sharp decline in investment and GDP (Katouzian, 1980). Meantime, public dissatisfaction with economic and non-economic policies motivated a revolutionary movement that led to the Islamic Republic’s establishment.

- **Period 4:1978-1988**

The Islamic Revolution and the Iran-Iraq war are two very important events in the modern history of Iran, which took place during this period. In this period, the annual GDP per capita growth rate and the inflation rate were -8 percent and 19 percent, respectively. Also, population increased by 4 percent annually. These events acted a turning point and slowed the positive trend of the previous period. They also had a huge negative effect on the Iranian economy through the present day (Acemoglu and Robinson, 2012)

## 3.1 The relationship between oil revenue and economic growth

Oil has been the backbone of Iranian economic development. For many years, oil earnings have provided a major part of government revenue. It is indeed the basic source of domestic investment (Pesaran, 1998). In 1973, oil prices and government's revenue increased because of Israeli-Arab war. In addition, after the Islamic Revolution the price of oil and government revenue continued to rise, but because of the United States sanction and Iran-Iraq war the revenue of government went down. As shown in Figure 2, value added of oil industry fluctuated significantly over time especially from 1979 to 1984, while non-oil GDP growth has had a smooth trend. The maximum quantity of the value added of oil growth rate is 128 percent, which was reported in 1982 (three years after Islamic Revolution) and the minimum quantity is -67 percent, which was reported in 1980. Indeed, based on the data when the price of oil increased dramatically, like 1973, 1979, and 2010, value added of oil sector decreased drastically. Moreover, as it is clear in Figure 2, GDP growth rate and value added of oil sector are highly correlated. The maximum quantity of real GDP growth rate was 23 percent, which was reported in 1982, the same year that the maximum quantity of oil's value added was reported, and the minimum quantity was -23 percent, which was reported in 1980, the same year that the minimum quantity of oil's value added was reported. It should be noted that in some years, like 1979, 1983, and 2010, the government's ability to sell oil and oil revenues decreased sharply due to the Islamic Revolution, the Iran-Iraq war and United Nations sanctions. Because of these events, we see the minimum quantity of GDP growth rate across these years. Furthermore, in 2015, after the Joint Comprehensive Plan of Action (JCPOA), the ability of the government to sell oil increased so that value added of oil went up significantly. Therefore, it can be inferred from the data that Iran's economy is highly dependent on the oil sector even in recent years. Additionally, the oil market has had significance effect on capital market of oil exporter countries like Canada, Iran, Qatar, and United Arab Emirates. (Ghaneei and Mahmoudi, 2021). In addition, in order to reduce dependency of Iran's economy on oil revenue some researchers recommend expanding renewable source of energy like wave energy, wind energy, solar energy.(Ghaneei and Mahmoudi, 2021)

*Figure 2.Oil and Non-Oil Real GDP Growth*

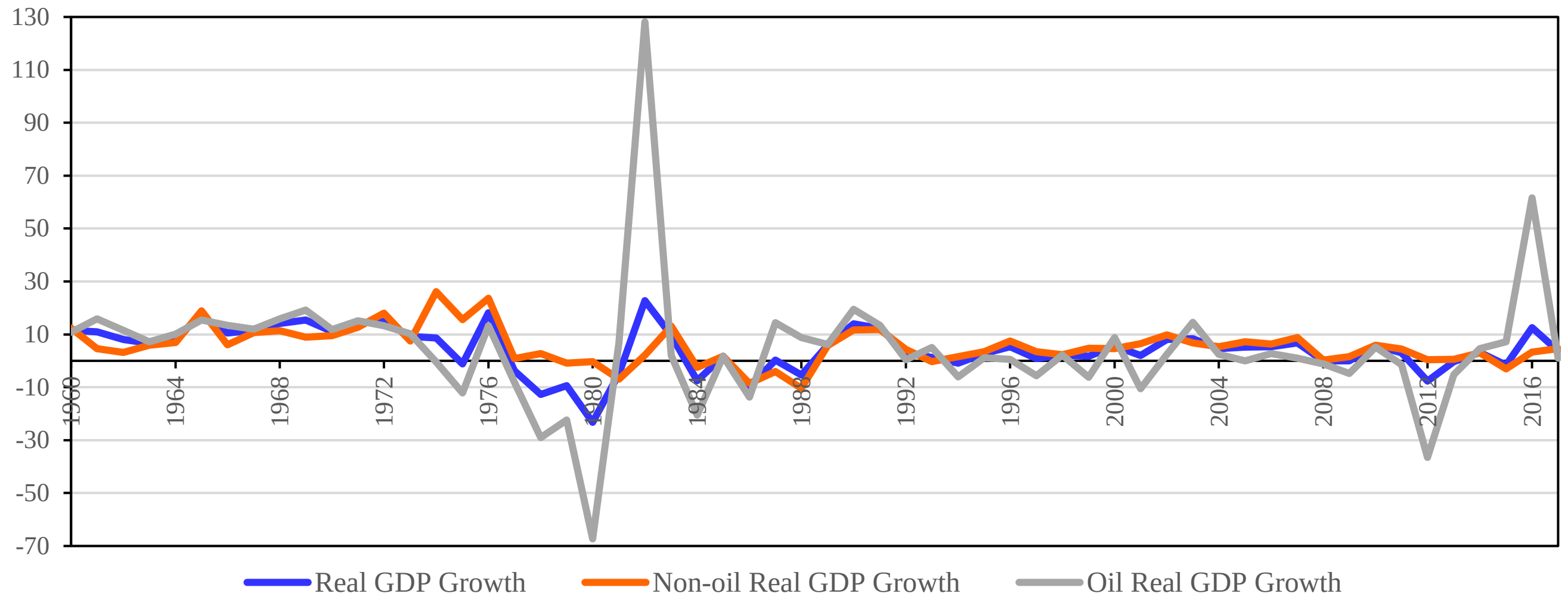

Source: Central Bank of Iran

## 3.2 Inflation rate

With the exception of the 1960s, Iranian economy suffered from a high rate of inflation for many years especially after the Islamic Revolution (Liu and Adedeji, 2000). Rising oil price is the main reason for the dramatic increase in inflation during 1973, 1979, 2008, and 2013, while the hyperinflation, which was reported around 1995, was associated with heavy foreign debt of government and subsequent dramatic decrease in national currency value. As it is clear in Figure 3, inflation rate and money supply are positively correlated. When the oil price goes up the government revenue goes up. This leads to government increasing its expenditure by increasing the money supply and following this, we see great inflation rate.

*Figure 3.Inflation and Money Supply Growth Rate*

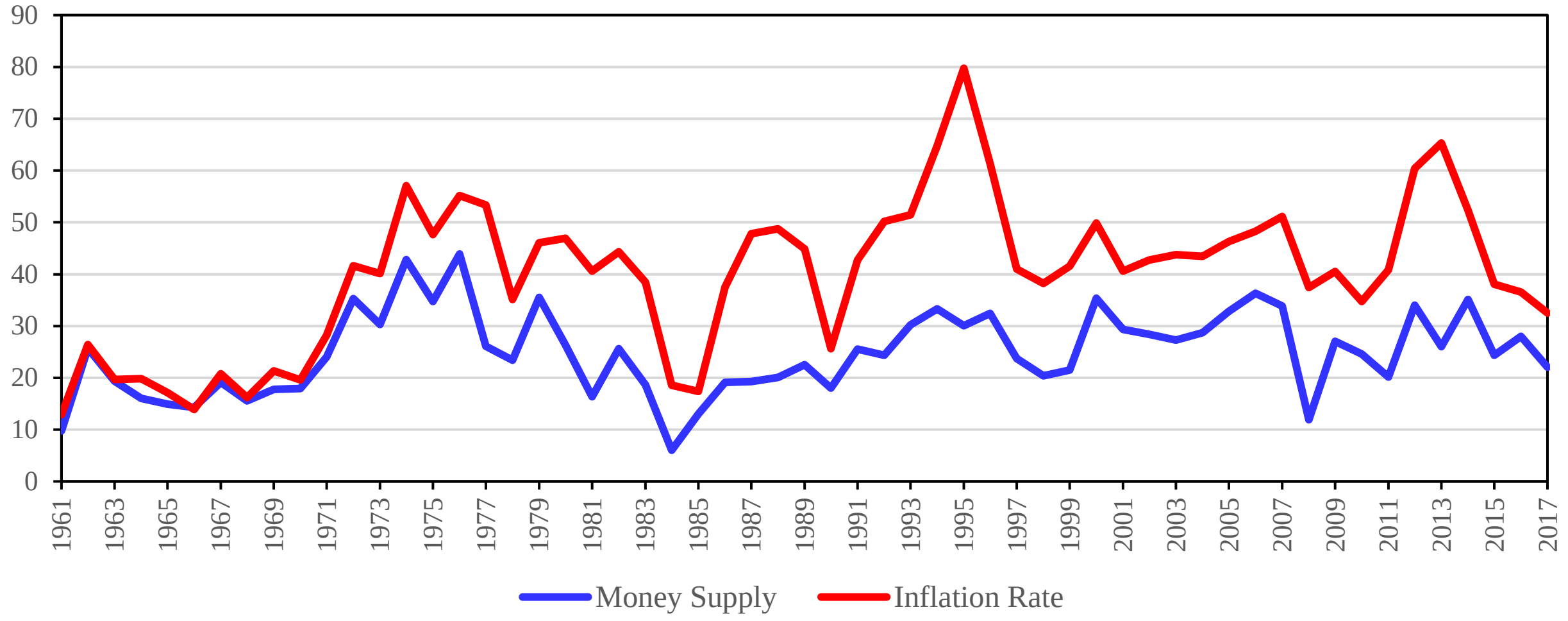

Source: World Bank, World Development Indicators 2018

## 3.3 Sectoral Shares of GDP in Iran

As described in Figure 4, the share of oil industry decreased in recent years, even though this sector still plays a critical role in the Iranian economy. The maximum share of oil sector of GDP was reported in 1973, due to oil crisis. After the Islamic Revolution this share decreased to 15 percent, but in following years it followed a stable trend. Moreover, as shown in this figure, share of industry sector rises from 10 percent to 25 percent. Furthermore, despite the stable trend of the agriculture sector, the share of services has had great fluctuation. The share of this part increases from 30 percent to 50 percent in recent years.

*Figure 4.The Ratio of Agriculture, Industry, Oil, and Services Sectors to GDP*

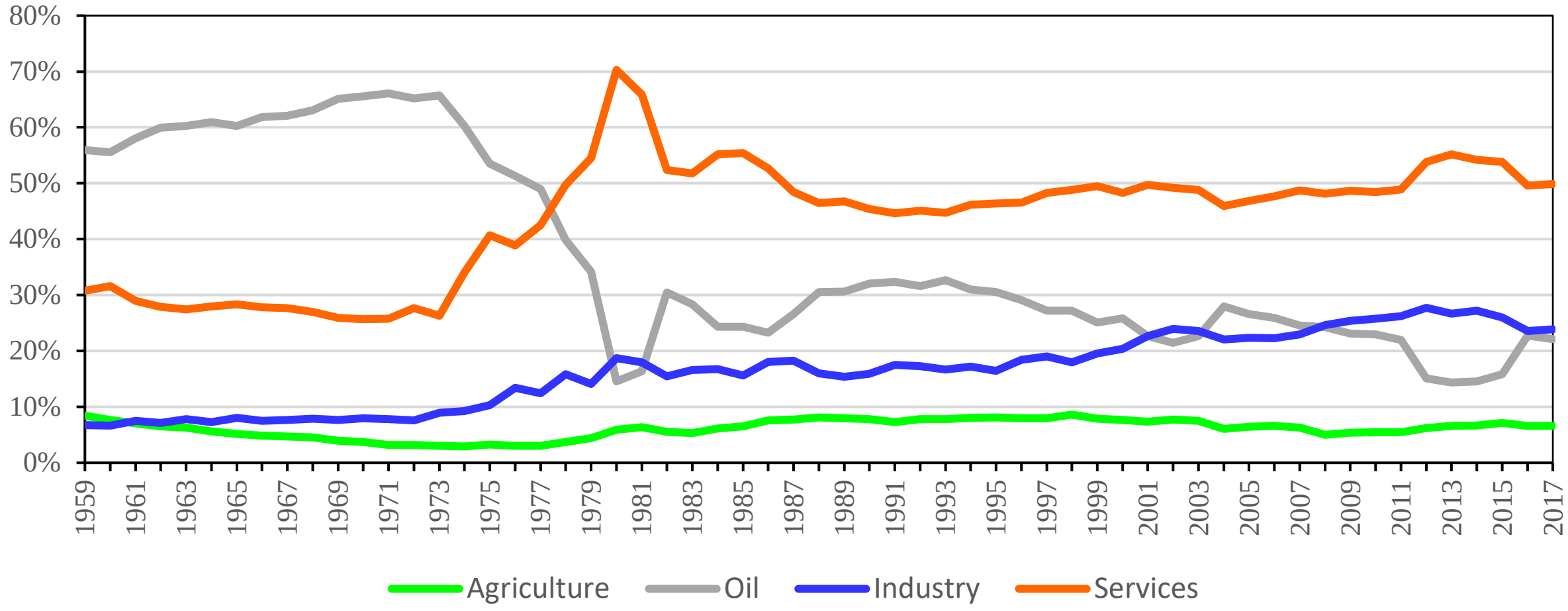

Source: Central Bank of Iran

Moreover, we analyzed the ratio of different components of the expenditure side of GDP. As illustrated in Figure 5, the share of net export has most fluctuation over time in comparison with other indicators. Especially between 1975 and 1986 because of revolution and Iran-Iraq war this index fluctuated dramatically. The share of investment fluctuated significantly, while the ratios of consumption and government expenditure have less variation.

*Figure 5.Shares of Consumption, Investment, Government purchases, and Net exports to Output*

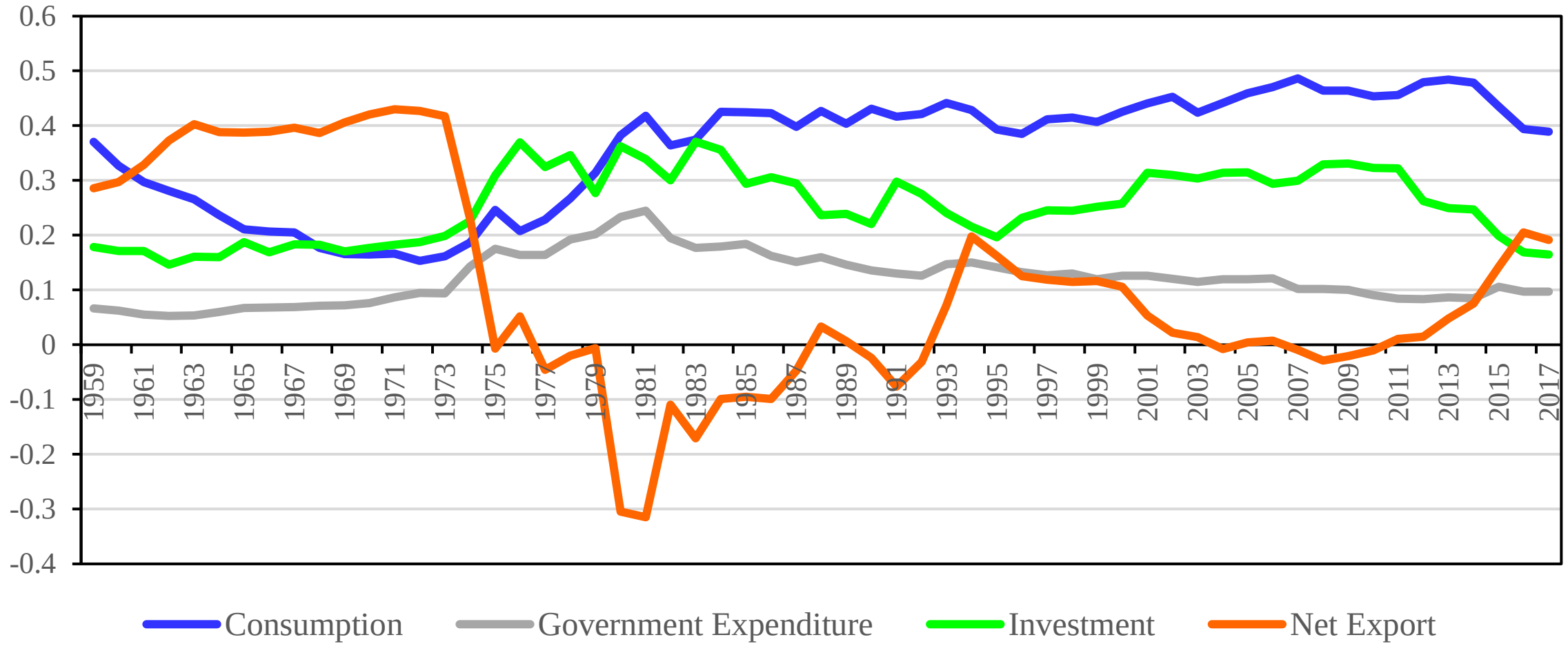

Source: Central Bank of Iran

## 4. Methodology

Growth accounting divides the growth in output of a country into two parts. The first part is the growth in output that can be attributed to growth in all factor inputs, holding technology constant. The second part is the growth that is solely due to an increase or decrease in technology. We use the following standard growth accounting framework, which was developed by (Solow, 1957) to analyze historical data of economic growth in Iran.

$$Y_t = A_t K_t^{\alpha} L_t^{1-\alpha} \quad (1)$$

$$Log\left(\frac{Y_{t+1}}{Y_t}\right) = Log\left(\frac{A_{t+1}}{A_t}\right) + \alpha Log\left(\frac{K_{t+1}}{K_t}\right) + (1-\alpha) Log\left(\frac{L_{t+1}}{L_t}\right) \quad (2)$$

Where Y, K, L, and A are output, capital stock, labor force and total factor productivity (TFP) which estimated as residual. Additionally, α represents the capital share. Based on The Penn World Table (version 9.0) data, the average share of labor for the Iranian economy is about 66 percent. Then, we calculate the share of capital easily by (1- Share of Labor), which equals to 34 percent. By multiplying these numbers by the capital and labor growth rate we compute contributions of capital and labor. Moreover, by subtracting the GDP growth rate from both capital and labor contributions we compute the contribution of technology. As seen in the Table 3, from 1960 to 2018, GDP grew by 94 percent; the contributions of capital, labor, and productivity are 33, 46, and 14 percent, respectively. However, each decade has different growth rate and contributions of labor, capital, and technology.

*Table 3.Growth Accounting for Iran*

| Period | Growth Rate | Contribution of Capital | Contribution of Labor | Contribution of TFP |
|---|---|---|---|---|
| **1960-2018** | 0.94 | 0.33 | 0.46 | 0.14 |
| **1960-1969** | 0.44 | 0.15 | 0.06 | 0.23 |
| **1970-1979** | 0.11 | 0.10 | 0.08 | -0.08 |
| **1980-1989** | 0.04 | -0.09 | 0.09 | 0.04 |
| **1990-1999** | 0.11 | 0.00 | 0.09 | 0.02 |
| **2000-2009** | 0.16 | 0.11 | 0.08 | -0.03 |
| **2010-2018** | 0.06 | -0.03 | 0.02 | 0.08 |

Source: The Penn World Table (version 9.0); World Bank, World Development Indicators 2018

Furthermore, as it is clear in Figure 6, contribution of capital plays a significant role in economic growth of Iran. Including the decades in which Iranian economy experienced depression for several years, the contribution of technology was significantly low. On the other hand, when the Iranian economy has growth for several years, the contribution of productivity is significantly high. These results are likely due to our estimation of total factor productivity as residual.

*Figure 6.Growth Accounting for Iran*

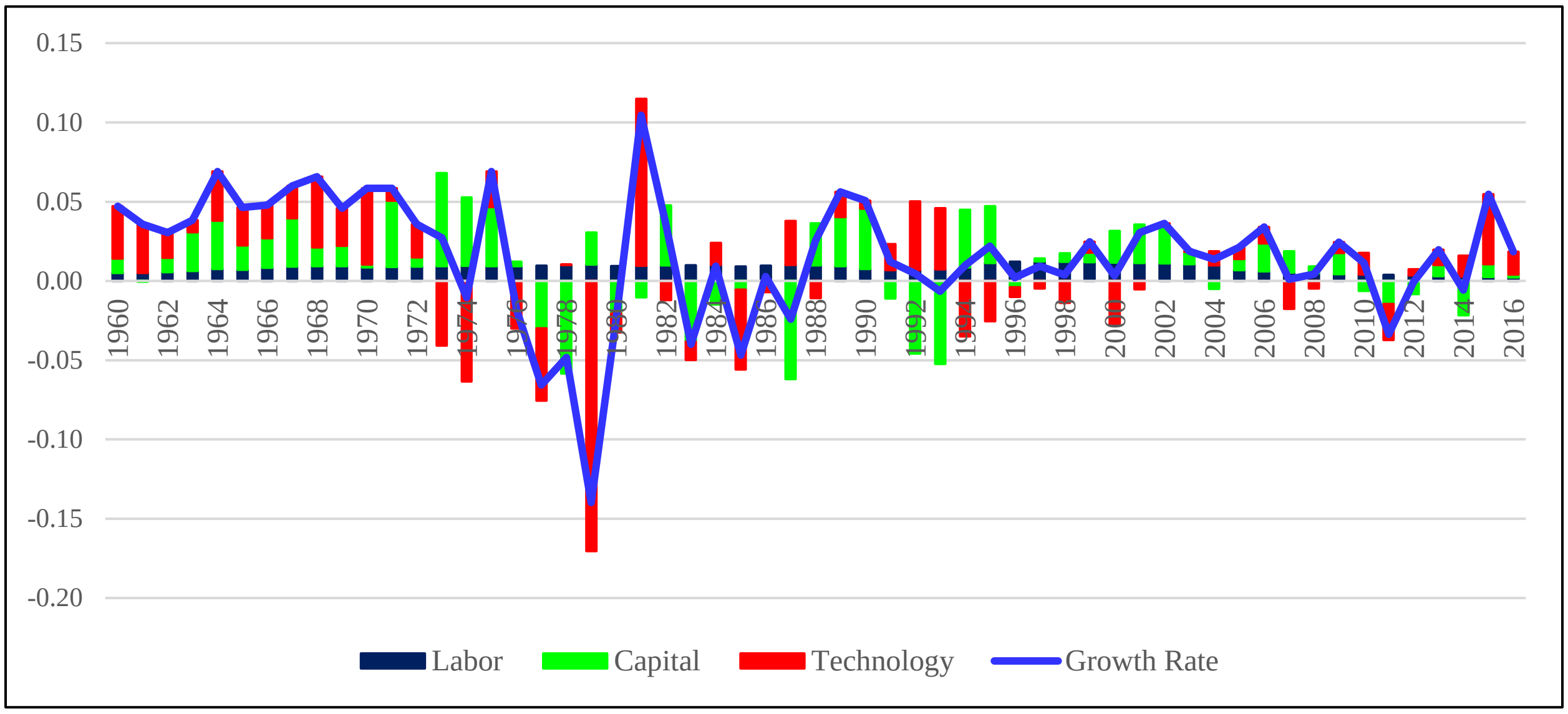

Source: The Penn World Table (version 9.0); World Bank, World Development Indicators 2018

## 4.1 Analyzing the steady state of capital for Iranian economy

In this section we compute the steady state of capital and analyze Iran's economy around the steady state based on the following model.

$$\sum_{t=0}^{\infty} \beta^t \frac{C_t^{1-\gamma}}{1-\gamma} \tag{3}$$

s.t

$$Y_t = A_t K_t^{\alpha} (l_t)^{1-\alpha} \tag{4}$$

$$C_t + I_t = Y_t \tag{5}$$

$$K_{t+1} = (1-\delta)K_t + I_t \quad (6)$$

$$C_t, K_{t+1} \geq 0 \quad (7)$$

In above system, $A_0 > 0$, $K_0 > 0$, and $L_0 > 0$ are given for all t. At time t, $C_t$ is consumption, $Y_t$ is output, $A_t$ is productivity, $K_t$ is physical capital, $L_t$ is labor, and $I_t$ is investment. The discount rate, β, satisfies $0 < \beta < 1$. The capital share, α, satisfies $0 < \alpha < 1$. The depreciation rate of capital, $\delta$, satisfies $0 < \delta < 1$. Relative risk-aversion, $\gamma$, satisfies $\gamma > 0 \; and \; \gamma \neq 1$. We assume that productivity grows at the constant rate a and that labor grows at the constant rate n.

Since productivity and labor growing at the constant rate, there is no steady state for above system. One way to resolve this issue is as follows;

We write production function like following equation;

$$Y_t = K_t^{\alpha}(A_t^{\frac{1}{1-\alpha}} L_t)^{1-\alpha} \quad (8)$$

By dividing both sides of equation 8 by $(A_t^{\frac{1}{1-\alpha}} L_t)$ we get equation 9.

$$y_t = k_t^{\alpha} \quad (9)$$

where $y_t = (\frac{Y_t}{A_t^{\frac{1}{1-\alpha}} L_t})$ is output for effective labor, and $k_t = (\frac{K_t}{A_t^{\frac{1}{1-\alpha}} L_t})$ is capital for effective labor.

We write the same process for other equations.

By Dividing both sides of equation 5 by $(A_t^{\frac{1}{1-\alpha}} L_t)$ we have;

$$c_t + i_t = y_t \quad (10)$$

Where $c_t = (\frac{C_t}{A_t^{\frac{1}{1-\alpha}} L_t})$, and $i_t = (\frac{I_t}{A_t^{\frac{1}{1-\alpha}} L_t})$

$$i_t = k_{t+1} - (1-\delta)k_t \tag{11}$$

Then, we write control variable as a function of state variable;

$$c_t = k_t^{\alpha} - k_{t+1} + (1-\delta)k_t \tag{12}$$

Now, this model has steady state and we write objective function as follows,

$$\sum_{t=0}^{\infty} \beta^t \frac{(k_t^{\alpha} - k_{t+1} + (1-\delta)k_t)^{1-\gamma}}{1-\gamma} \tag{13}$$

We want to find derivative with respect to $k_{t+1}$. $k_{t+1}$ appears in two periods t and t+1, so that the derivative of objective function with respect to $k_{t+1}$ is calculated as follows;

$$\ldots + \beta^t \frac{(k_t^{\alpha} - k_{t+1} + (1-\delta)k_t)^{1-\gamma}}{1-\gamma} + \beta^{t+1} \frac{(k_{t+1}^{\alpha} - k_{t+2} + (1-\delta)k_{t+1})^{1-\gamma}}{1-\gamma} + \ldots \tag{14}$$

F.O.C w.r.t $k_{t+1}$:

$$-\beta^t(k_t^{\alpha} - k_{t+1} + (1-\delta)k_t)^{-\gamma} + \beta^{t+1}(k_{t+1}^{\alpha} - k_{t+2} + (1-\delta)k_{t+1})^{-\gamma}\left(\alpha k_{t+1}^{\alpha-1} + (1-\delta)\right) = 0 \tag{15}$$

By Dividing equation 15 by $\beta^t$ we get equation 14.

$$(c_t)^{-\gamma} = \beta(c_{t+1})^{-\gamma}\left(\alpha k_{t+1}^{\alpha-1} + (1-\delta)\right) \tag{16}$$

At balance growth path $\frac{c_{t+1}}{c_t} = \frac{y_{t+1}}{y_t} = 1 + g$, also at the steady state we have $\bar{k}$, then we write equation 16 as follows,

$$\left(\frac{c_{t+1}}{c_t}\right)^{\gamma} = \beta\left(\alpha\bar{k}^{\alpha-1} + (1-\delta)\right) \tag{17}$$

$$(1+g)^{\gamma} = \beta\left(\alpha\bar{k}^{\alpha-1} + (1-\delta)\right) \tag{18}$$

$$\frac{(1+g)^{\gamma} - \beta(1-\delta)}{\alpha\beta} = \bar{k}^{\alpha-1} \tag{19}$$

Therefore, the steady state of effective capital per person is equal to:

$$\bar{k} = \left(\frac{\alpha\beta}{(1+g)^{\gamma} - \beta(1-\delta)}\right)^{\frac{1}{1-\alpha}} \tag{20}$$

Moreover, we calculate (1+g) using production function as follows;

$$Y_t = A_t K_t^{\alpha} (L_t)^{1-\alpha} \tag{21}$$

$$\frac{Y_{t+1}}{Y_t} = \left(\frac{A_{t+1}}{A_t}\right)\left(\frac{K_{t+1}}{K_t}\right)^{\alpha}\left(\frac{L_{t+1}}{L_t}\right)^{1-\alpha} \tag{22}$$

By considering balance growth path relations and the growth rate of labor and productivity, we write equation 22 as follows,

$$(1+g) = (1+a)(1+n)^{1-\alpha}(1+g)^{\alpha} \tag{23}$$

$$(1+g)^{1-\alpha} = (1+a)(1+n)^{1-\alpha} \tag{24}$$

$$(1+g) = (1+a)^{\frac{1}{1-\alpha}}(1+n) \tag{25}$$

We write objective function and balance growth path again for calculating $\frac{K_t}{Y_t}, and\ \frac{I_t}{Y_t}$.

$$C_t = A_t K_t^{\alpha} L_t^{1-\alpha} - K_{t+1} + (1-\delta)K_t \tag{26}$$

The derivative of objective function with respect to $k_{t+1}$ is as follows;

$$\ldots + \beta^t \frac{(A_t K_t^{\alpha} L_t^{1-\alpha} - K_{t+1} + (1-\delta)K_t)^{1-\gamma}}{1-\gamma} + \beta^{t+1} \frac{(A_{t+1} K_{t+1}^{\alpha} L_{t+1}^{1-\alpha} - K_{t+2} + (1-\delta)K_{t+1})^{1-\gamma}}{1-\gamma} + \ldots \tag{27}$$

F.O.C w.r.t $K_{t+1}$:

$$(C_t)^{-\gamma} = \beta(C_{t+1})^{-\gamma}\left(\alpha A_{t+1} K_{t+1}^{\alpha} L_{t+1}^{1-\alpha} + (1-\delta)\right) \tag{28}$$

Where, $A_{t+1} K_{t+1}^{\alpha} L_{t+1}^{1-\alpha} = \frac{Y_{t+1}}{K_{t+1}}$, and based on balance growth path $\frac{C_{t+1}}{C} = \frac{Y_{t+1}}{Y_t} = \frac{K_{t+1}}{K_t} = 1+g$

Hence, we write equation 28 as follows;

$$(1+g)^{\gamma} = \beta\left(\alpha \frac{Y_{t+1}}{K_{t+1}} + (1-\delta)\right) \tag{29}$$

$$\frac{(1+g)^{\gamma} - \beta(1-\delta)}{\alpha\beta} = \frac{Y_{t+1}}{K_{t+1}} \tag{30}$$

$$\frac{K_{t+1}}{Y_{t+1}} = \frac{\alpha\beta}{(1+g)^{\gamma} - \beta(1-\delta)} \tag{31}$$

Therefore, we compute $\frac{K_t}{Y_t}$ as follows,

$$\frac{K_{t+1}}{Y_{t+1}}\frac{Y_{t+1}}{K_{t+1}}\frac{K_t}{Y_t}=\frac{\alpha\beta}{(1+g)^{\gamma}-\beta(1-\delta)}\frac{Y_{t+1}}{K_{t+1}}\frac{K_t}{Y_t} \tag{32}$$

$$\frac{K_t}{Y_t}=\frac{\alpha\beta}{(1+g)^{\gamma}-\beta(1-\delta)} \tag{33}$$

Also, we do the following process for computing $\frac{I_t}{Y_t}$.

$$I_t = K_{t+1} - (1-\delta)K_t \tag{34}$$

$$\frac{I_t}{Y_t}= \frac{K_{t+1}}{Y_t}-(1-\delta)\frac{K_t}{Y_t} \tag{35}$$

To compute $\frac{K_{t+1}}{Y_t}$, we use equation 31 and rewrite it as follows,

$$\frac{K_{t+1}}{Y_{t+1}}\frac{Y_{t+1}}{Y_t}=\frac{\alpha\beta}{(1+g)^{\gamma}-\beta(1-\delta)}\frac{Y_{t+1}}{Y_t} \tag{36}$$

$$\frac{K_{t+1}}{Y_t}=\frac{\alpha\beta}{(1+g)^{\gamma}-\beta(1-\delta)}\,(1+g) \tag{37}$$

Therefore, $\frac{I_t}{Y_t}$ equals to:

$$\frac{I_t}{Y_t}= \frac{\alpha\beta}{(1+g)^{\gamma}-\beta(1-\delta)}\,(1+g)- (1-\delta)\frac{\alpha\beta}{(1+g)^{\gamma}-\beta(1-\delta)}=\frac{\alpha\beta}{(1+g)^{\gamma}-\beta(1-\delta)}(g+\delta) \tag{38}$$

We compute different formula that enable us to calculate parameters, and steady state. In following, the process of calibrating parameters for Iranian economy is explained step by step.

- In the first step, we define steady state period. Based on the data, the performance of the Iranian economy between 1996 to 2005 satisfies the steady state condition. During this period the government was politically stabilized and it did not face with any significant internal or external challenges. Moreover, as it is clear in Figure 7, during this period output and consumption were likely growing at the same rate, which means $\frac{Y_{t+1}}{Y_t} = \frac{C_{t+1}}{C_t}$ =1+g.

*Figure 7.Growth Rate of Output and Consumption*

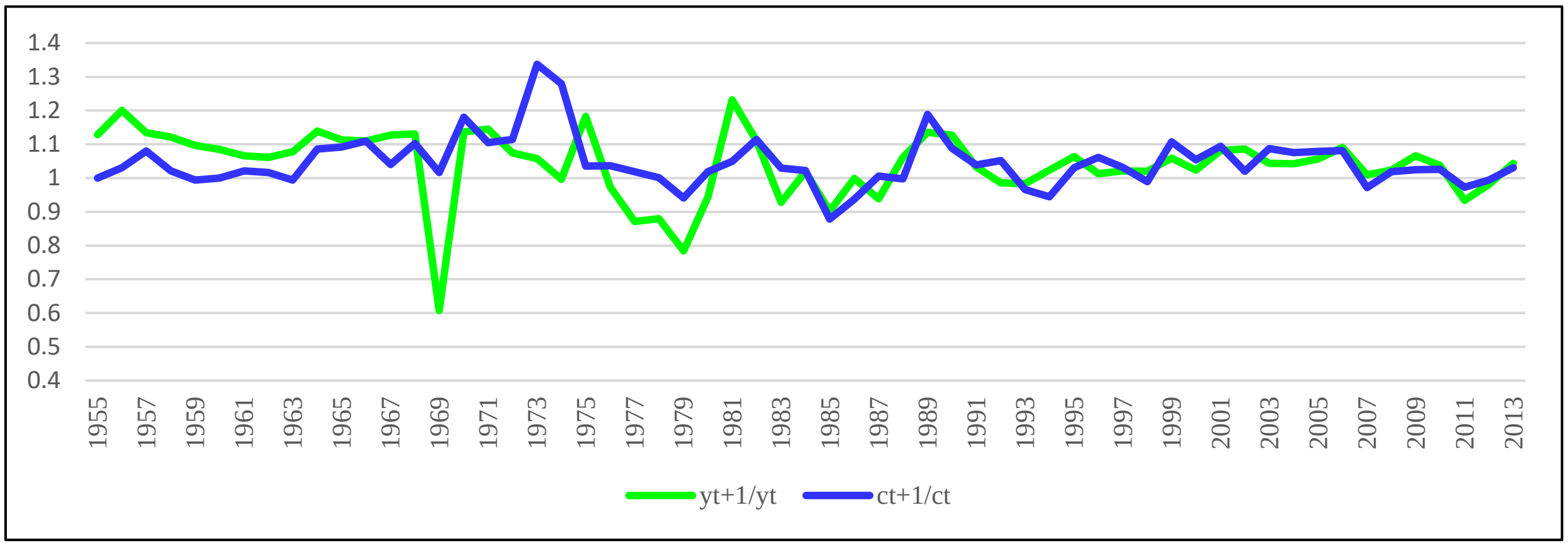

Source: The Penn World Table (version 9.0)

In the second step, we calibrate the parameters for this period as follows.

- Based on the Penn World Table (version 9.0) data, the average capital share of Iran's economy,α, is 34 percent. Also, growth rate of productivity, a, and growth rate of labor, n, are .04 percent and 2 percent, respectively. By plugging the quantities of n and a into the equation 25 we calculate (1+g), which equals to 1.02.
- Based on The Penn World Table (version 9.0) data, the average depreciation rate of capital for Iran's economy is 5 percent.
- For calibrating other parameters, β and γ, we use the following procedure:
    - First, we calculate the average of $\frac{I_t}{Y_t}$ for the steady state period. This term equals to 0.21.

      By Plugging the quantities for parameters in equation 25, we have the following formula that show us the relation between β and γ.

$$0.21 = \frac{.33\beta}{(1.02)^{\gamma} - \beta(.95)}(.07) \qquad (39)$$

- Second, we calculate the average of $\frac{K_t}{Y_t}$ for the steady state period. This term equals to 2.63. By Plugging the quantities for parameters in equation 25, we have another formula that show us the relation between β and γ.

$$2.63 = \frac{.33\beta}{(1.02)^{\gamma} - \beta(.95)} \tag{40}$$

- Third, we put different quantities of β and γ in above equations and choose the ones that match the steady state values of $\frac{I_t}{Y_t}$ and $\frac{K_t}{Y_t}$ to real data. We do the matching process through 3 scenarios. As it is clear in Table 4, the best quantities of β and γ which give us the closest quantities of $\frac{K_t}{Y_t}$ and $\frac{I_t}{Y_t}$ to the real data are 0.93 and 0.4, respectively.
- Finally, by plugging the quantities of different parameters into the equation 20 we compute steady state of effective capital, which is 4.27 for Iran's economy.

*Table 4.Calibrating the Parameters*

| | β | γ | $\boldsymbol{\frac{K_t}{Y_t}}$ | $\boldsymbol{\frac{I_t}{Y_t}}$ | $\boldsymbol{\bar{k}}$ |
|---|---|---|---|---|---|
| **Scenario 1** | 0.97 | 1.8 | 2.78 | 0.20 | 4.72 |
| **Scenario 2** | 0.94 | 0.20 | 2.8 | 0.20 | 4.8 |
| **Scenario 3** | **0.93** | **0.4** | **2.61** | **0.201** | **4.27** |

Source: Researcher Computations

## 5. Conclusion

Although GDP per capital increased by 2 percent annually from 1950 to 2018, this indicator has had a huge fluctuation over time. Before the Islamic Revolution Iran experienced high rate of economic growth and low rate of inflation especially from 1961 to 1971. However, after the Islamic Revolution we see a great fluctuation in growth rate and high level of inflation rate. Historical data shows a close relationship between economic growth of Iran and oil revenue. Even though the shares of other sectors like industry and services in GDP have increased over the time, oil sector still plays a key role in economic growth of Iran. Furthermore, growth accounting framework shows contribution of capital affects Iranian economic growth rate significantly and the steady state of effective capital is 4.27 for Iran's economy.